\documentclass[preprint,superscriptaddress,showkeys,nofootinbib,showpacs,
   preprintnumbers,amsmath,amssymb,aps,floatfix]{revtex4-1}
\usepackage{graphicx}
\usepackage{epsfig}
\usepackage{amsfonts}
\usepackage{amsmath}
\usepackage{amssymb}
\usepackage[sort&compress]{natbib}
\usepackage{dcolumn}
\usepackage{multirow}
\usepackage{bigstrut}
\usepackage{type1cm}
\usepackage{placeins}
\usepackage{rotating}
\usepackage{slashed}
\usepackage[caption=false]{subfig}

\newcommand{\al}{\alpha}
\newcommand{\bt}{\beta}

\newcommand{\simu}{\sigma^{\mu\nu}}

\newcommand{\vL}{\ensuremath{\mathcal{L}}}

\newcommand{\MQCD}{M_{\mathrm{QCD}}}

\newcommand{\Or}{\mathcal O}

\newcommand{\vp}{\varphi}
\newcommand{\sq}{^{2}}
\newcommand{\ga}{\gamma}

\newcommand{\dslash}[1]{#1 \llap{/\kern-0.5pt}}
\newcommand{\Dslash}[1]{#1 \llap{/\kern+1.2pt}}
\newcommand{\DDslash}[1]{#1 \llap{/\kern+2.3pt}}
\newcommand{\dslashh}[1]{#1 \llap{/\kern+1pt}}

\newcommand{\bea}{\begin{eqnarray}}
\newcommand{\eea}{\end{eqnarray}}
\newcommand{\bma}{\begin{pmatrix}}
\newcommand{\ema}{\end{pmatrix}}
\newcommand{\nn}{\nonumber}

\begin{document}

\preprint{\today}

\title{$T$ violation in radiative $\beta$ decay and electric dipole moments}

\author{W. G. Dekens and K. K. Vos\\ \textit{Van Swinderen Institute, Faculty of Mathematics and Natural Sciences, \\ University of Groningen, Nijenborgh 4, \\
                   NL-9747 AG Groningen, The Netherlands}}

\date{\today}
\vspace{3em}

\begin{abstract}
In radiative $\beta$ decay, $T$ violation can be studied through a spin-independent $T$-odd correlation. 
We consider contributions to this correlation by beyond the standard model (BSM) sources of $T$-violation, arising above the electroweak scale. At the same time such sources, parametrized by dimension-6 operators, can induce electric dipole moments (EDMs). As a consequence, the manifestations of the $T$-odd BSM physics in radiative $\bt$ decay and EDMs are not independent. Here we exploit this connection to show that current EDM bounds already strongly constrain the spin-independent $T$-odd correlation in radiative $\bt$ decay.

%$T$ violation generated by beyond the standard model (BSM) sources above the electroweak scale. The induced $T$ violation in radiative $\bt$ decay is not independent from the contributions to electric dipole moment (EDM), and we conclude that EDM bounds strongly constrain the $T$-odd correlation in radiative $\beta$ decay. 

\end{abstract}
\maketitle
\section{Introduction}
The Standard Model of particle physics (SM) cannot account for the baryon asymmetry of the universe \cite{Riotto:1999yt,Kuzmin,Sak67}, and additional sources of $CP$ violation might be expected to arise beyond the SM (BSM). Searches for additional time-reversal ($T$) violation, and equivalently $CP$ violation, are, therefore, promising probes of BSM physics.
Especially interesting are observables with a very low SM background, such as the electric dipole moments (EDMs) of hadrons, nuclei, atoms, and molecules. 

In $\bt$ decay, $T$ violation is probed by the triple-correlation coefficients $D$ and $R$ \cite{Jac57}. However, these observables are not independent from EDM measurements. In fact, the stringent neutron EDM limit bounds $D$ more than an order of magnitude better than current $\beta$-decay experiments \cite{Ng12, Her05, Mum11,Chu12}. Molecular and atomic EDMs constrain scalar and tensor electron-nucleon couplings, which leads to strong constraints on the $R$ coefficient. These constraints are several order of magnitude better than the current best $\beta$-decay bounds \cite{Vos15, Khr91, Koz09, Hub03}. 

In radiative decays it is possible to study spin-independent $T$-odd triple-correlations \cite{Bra01,Gar12,Gar13}, which are not present in $\beta$ decay. In this paper, we consider such a correlation in radiative $\beta$ decay generated by high-energy BSM sources of CP violation. As in $\beta$ decay, we find that this $T$-odd correlation and EDMs are connected, which allows EDM bounds to strongly constrain the spin-independent $T$-odd correlation.

We work in an effective field theory (EFT) framework in which dimension-6 operators parametrize the new sources of $CP$ violation. We first discuss these operators. Then we consider their contributions  to radiative $\beta$ decay in Section \ref{sec:radbeta}, while discussing the contribution of these operators to the EDM in section \ref{sec:EDM}. Finally, we give the current EDM bounds on these operators and end with a brief discussion.

\section{Formalism}\label{sec:formalism}
We consider the effects of new $T$-violating physics on the correlation $K \vec{p}_\nu\cdot (\vec{p}_e\times \vec{k})$, where $\vec{k}$ is the photon momentum, and neglect the small $T$-violating SM contributions generated by the $CP$-odd phase of the CKM matrix and the QCD $\theta$-term \cite{Her97}. Besides these true SM $T$-odd sources, there are also electromagnetic final-state interactions (FSI) that mimic $T$-violation and that also contribute to the triple-correlation (similar FSI contribute to $D$ and $R$). These FSI have been studied for the neutron, $^{19}$Ne, and $^{35}$Ar \cite{Gar12, Gar13a, Gar13}, and contribute to the $T$-odd asymmetry at $\mathcal{O}(10^{-3}) - \mathcal{O}(10^{-5})$, depending on the detectable photon energy and the used isotope. 

The effects of new $T$-violating physics, arising at a high scale $\Lambda$, can be studied in an EFT framework. At low energies, the new physics is effectively described by higher-dimensional operators. We consider dimension-six operators, for which the complete set of gauge-invariant operators has been derived in Ref.~\cite{Grz10}. We divide the operators relevant for radiative $\beta$ decay into two groups.
%while higher-dimensional operators are suppressed by powers of the new physics scale $\Lambda$.  

($i$) The first group consists of four-fermion operators that also contribute to $\beta$ decay \cite{Grz10,Cir10}. The relevant part of the effective $\beta$-decay Lagrangian is \cite{Her01}
\begin{align}\label{eq:lageff}
\mathcal{L}^{(\textrm{eff})}_{S,P,T} & = \frac{-4G_{F}}{\sqrt{2}} \sum_{\epsilon,\delta = L, R} \left\{A_{\epsilon\delta}\; \bar{e} \nu^{\epsilon}_e\cdot\bar{u}d_{\delta} + \alpha_{\epsilon}\: \bar{e}\frac{\sigma^{\mu\nu}}{\sqrt{2}}\nu^{\epsilon}_e\cdot \bar{u} \frac{\sigma_{\mu\nu}}{\sqrt{2}} d_{\epsilon} \right\}\ +\text{h.c.},
\end{align}
where we have set $V_{ud}=1$ for convenience. $G_{F}$ is the Fermi coupling constant and we sum over the chirality ($L$, $R$) of the final states. These four-fermion operators modify the $V-A$ coupling of the SM, by generating scalar/pseudoscalar ($A$) and tensor ($\alpha$) couplings \cite{Lee56,Jac57}. Besides contributing to $\beta$ decay, the operators in Eq.~\eqref{eq:lageff} also contribute to radiative $\beta$ decay after being dressed with bremsstrahlung photons. 

($ii$) The second group of $T$-violating operators is given in Table \ref{tab:ops}\footnote{In principle, the operator $Q_{eW} = (\bar{l} \sigma^{\mu\nu}e)\tau^I \varphi W_{\mu\nu}^I$ also contributes to radiative $\beta$ decay, however, it does not contribute to $K$ at leading recoil order. }. At the scale of new physics, $\Lambda$, the relevant terms for radiative $\beta$ decay, are
\bea \label{HighELag}
\vL_6& =&   C_{\vp\tilde WB}(\Lambda)\frac{g c_w v\sq }{2} i\varepsilon^{\mu\nu\al\bt}W^+_\mu W^-_\nu F_{\al\bt}+C_{\vp ud}(\Lambda) \frac{v\sq g}{2\sqrt{2}} \bar u_R \ga^\mu d_R W^+_\mu\nn\\
&& + 2 v  C_{uW}(\Lambda) \, (\bar d_L  \simu \overset{\leftrightarrow}{D}_\nu u_R )W^-_{\mu} +2v  C_{dW}(\Lambda) \, (\bar u_L  \simu \overset{\leftrightarrow}{D}_\nu d_R) W^+_{\mu} \nn\\
&& +  \text{h.c.}+\dots,
\eea
where $v\approx 246 \, \text{GeV}$ is the vacuum expectation value of the Higgs field $\langle\vp\rangle = \frac{1}{\sqrt{2}} v$, the photon field is denoted by $A_\mu$ and $s_w=\sin\theta_w$ is the sine of the Weinberg angle ($c_w=\cos\theta_w$). The covariant derivative $D_\mu = \partial_\mu - i s_w g q_f A_\nu$, where $q_f$ is the charge of the fermion. $C_X$ is the coupling constant associated with the operator $Q_X$ defined in Table~\ref{tab:ops}.

\begin{figure}[h]
	\centering
\subfloat[$Q_{\varphi ud}$]{\label{fig:1}\includegraphics[width=0.25\textwidth]{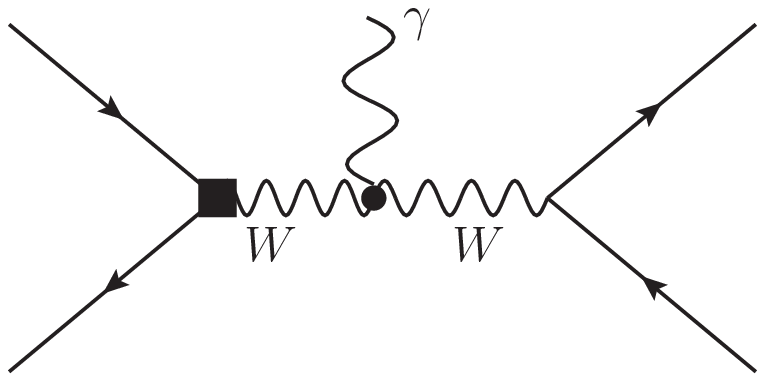}}
\subfloat[$Q_{uW, dW}$]{\label{fig:2}\includegraphics[width=0.25\textwidth]{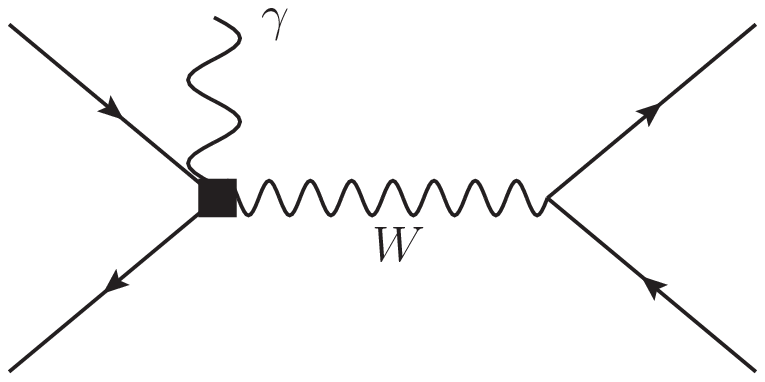}}	%
\subfloat[$Q_{\varphi \tilde{W} B}$]{\label{fig:3} \includegraphics[width=0.25\textwidth]{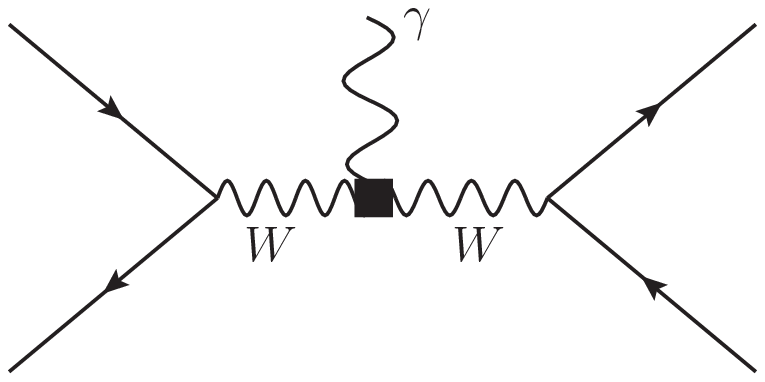}}
\caption{The relevant $T$-violating effective interactions for radiative $\beta$ decay. The box indicates one of the BSM interactions of Table \ref{tab:ops}, the dot indicates the SM coupling of two $W$ bosons and a photon.}
	\label{fig:new coupling}
\end{figure}

Fig.~\ref{fig:new coupling} shows how these operators contribute to radiative $\bt$ decay. At low energies, $\mu\approx 1\, \text{GeV}$, after integrating out the $W^\pm$ boson, we obtain
\bea
\vL^{\text{eff}}_6 &=& 
-\frac{8i c_w}{g v\sq}V_{ud}\,\text{Re}\,C_{\vp\tilde WB}(\Lambda)\, \varepsilon^{\mu\nu\al\bt}(\bar u_L\ga_\mu d_L)(\bar e_L\ga_\nu \nu_L)F_{\al\bt} +\frac{1}{M_W\sq}C_{\vp ud}(\Lambda)(\bar u_R\ga_\mu d_R)\Gamma^{\mu\nu}(\bar e_L\ga_\nu \nu_L)\nn\\
&& -\frac{8is_w}{\sqrt{2}v}\eta_{qW} C_{Wu}^*(\Lambda)\,(\bar u_R \simu d_L )(\bar e_L\ga_\mu \nu_L)A_\nu -\frac{8is_w}{\sqrt{2}v}\eta_{qW} C_{Wd}(\Lambda)\,(\bar u_L \simu d_R )(\bar e_L\ga_\mu \nu_L)A_\nu  \nn\\&& +\text{h.c.} + \dots,
\label{LowELag}\eea
where $\Gamma^{\mu\nu} = g^{\mu\nu}D\sq -D^\nu D^\mu -i g s_w F^{\mu\nu}$, whose leading contribution to $K$ arises from Fig.\ \ref{fig:1}. Furthermore, $\eta_{qW} = \big(\frac{\al_s(\Lambda)}{\al_s(m_t)}\big)^{4/21}\big(\frac{\al_s(m_t)}{\al_s(m_b)}\big)^{4/23}\big(\frac{\al_s(m_b)}{\al_s(m_c)}\big)^{4/25}\big(\frac{\al_s(m_c)}{\al_s(\mu)}\big)^{4/27}$  is a running factor (numerically, $\eta_{qW} = 0.39\, (0.33)$ for $\Lambda = 1\,(10)\,\text{TeV})$, arising from the QCD renormalization of the $Q_{qW}$ operators \cite{Ciuchini:1993fk,Degrassi:2005zd}.  
The dots represent terms which are necessary to maintain gauge invariance, but that do not contribute to $K$ at leading recoil order. 

The first term in Eq.~\eqref{LowELag} is similar to the interaction studied in Eq.~(2) of Ref.~\cite{Gar13}. Although we find that such a term is not $T$-violating when it arises from a pseudo-Chern-Simons term ({\it i.e.} Eq.~(1) in Ref.~\cite{Gar13}), it is clear that it can be generated by BSM physics such as $Q_{\vp\tilde WB}$. 

%However, we disagree that such a term can violate $T$ symmetry if it arises from a pseudo-Chern-Simons term.   We find that such a term is not $T$-violating when it 

\begin{table}
	\centering
\begin{tabular}{  c || c }
  \hline	\hline		
 $Q_{\varphi ud}$ & $i (\tilde{\varphi}^\dagger D_\mu \varphi)(\bar{u} \gamma^\mu d)$  \\
  $Q_{\varphi \tilde{W}B} $& $\varphi^\dagger \tau^I \varphi \tilde{W}_{\mu\nu}^I B^{\mu\nu}$ \\
		$Q_{uW}$	& $(\bar{q} \sigma^{\mu\nu}\tau^I \tilde\varphi \,u)W_{\mu\nu}^I$  \\
			$Q_{dW}$	& $(\bar{q} \sigma^{\mu\nu}\tau^I \varphi \,d)W_{\mu\nu}^I$  \\
  \hline  \hline
\end{tabular}
	\caption{\footnotesize{Dimension-six operators that contribute to $T$-violating radiative $\bt$ decay. Here $\tau^I$ are the Pauli matrices, $\vp$ is the Higgs doublet  and $\tilde\vp = i\tau_2 \vp^*$. Furthermore, $D_\mu=\partial_\mu - i \frac{g}{2}\tau^I W^I_{\mu}-i\frac{g'}{2}B_{\mu} $ is the covariant derivative of the Higgs doublet, while $W^I_{\mu\nu} = \partial_\mu W_\nu^I-\partial_\nu W_\mu^I +g\varepsilon^{IJK}W_\mu^JW_\nu^K$ and $B_{\mu\nu} = \partial_\mu B_\nu-\partial_\nu B_\mu$ are the field strengths of the $SU(2)$ and $U(1)_Y$ gauge fields respectively. Finally, the duals of the field strengths are $\tilde X_{\mu\nu} = \varepsilon_{\mu\nu\al\bt}X^{\al\bt}$, where $\varepsilon^{0123}=+1$. }}
	\label{tab:ops}
\end{table}

\section{$T$-violating radiative $\beta$ decay}\label{sec:radbeta}
The new sources of $T$ violation contribute to the radiative $\beta$ decay rate 
\begin{equation}
d\Gamma = 32e^2 G_F^2  M_n M_p d\Gamma_0\left[ K \vec{p}_\nu\cdot(\vec{p}_e\times \vec{k})+ \cdots \right] \ ,
\end{equation}
where $M_{n,p}$ are the neutron and proton masses and $d\Gamma_0$ contains the integral over the phase space. The dots represent higher-order recoil terms as well as $T$-even terms that are present in the SM \cite{Ber04}.
The $K$ coefficient can be inferred from the asymmetry \cite{Gar12, Gar13}
\begin{equation}
\mathcal{A} = \frac{\Gamma^ +  - \Gamma^-}{\Gamma^ +  + \Gamma^-} \ , 
\end{equation}
where $\Gamma^+$ corresponds to the $\phi_\nu$ range $[0, \pi]$ and $\Gamma^-$ to the $\phi_\nu$ range $[\pi, 2\pi]$ if $\vec{p}_e$ is in the $\hat{z}$ direction, such that $\vec{k}$ and $\vec{p}_e$ fix the $\hat{z}-\hat{x}$ plane \cite{Gar12, Gar13}. The asymmetry depends on the $Q$-value of the interaction and on the threshold energy of the photon detector $\omega_{\textrm{min}}$, typically in the range of MeV. The asymmetry grows with increasing $\omega_{\textrm{min}}$, following Ref.~\cite{Gar13} we evaluate the integrals at $\omega_{\textrm{min}}=0.3$MeV.     
We discuss the form of $K$ and the contribution to $\mathcal{A}$ for the two groups of operators. 

($i$) The first group contributes to radiative $\beta$ decay after being dressed with bremsstrahlung photons \cite{Gar13}. For neutron decay,
\begin{equation}\label{eq:Ki}
K =   2 \frac{1}{M_p}\frac{1}{k\cdot p_e} \textrm{Im} \left[g_T\alpha_L( g_S^* A_L^{*} + g_P^* {A'}_L^{*}) - g_T\alpha_R (g_S^* {A}_R^{*} + g_P^* {A'}_R^{*})\right] \ ,
\end{equation}
where $A_L\equiv A_{LL} + A_{LR}$, $A_R\equiv A_{RR} + A_{RL}$, $A'_L\equiv A_{LL} - A_{LR}$ and $A'_R\equiv A_{RR} - A_{RL}$. The couplings $g_{\Lambda}$ are defined by $\left\langle p| \bar{u}\Gamma d|n\right\rangle = g_{\Gamma} \bar{p}\Gamma n$, with $\Gamma = 1, \gamma_5, \gamma_\mu, \gamma_\mu\gamma_5, \simu$. For $\omega_{\textrm{min}}=0.3$ MeV, the asymmetry is
\begin{equation}\label{LeeYangAsym}
\mathcal{A} = 2.1\times 10^{-5} \textrm{Im} \left[g_T\alpha_L( g_S^* A_L^{*} + g_P^* {A'}_L^{*}) - g_T\alpha_R (g_S^* {A}_R^{*} + g_P^* {A'}_R^{*})\right] \ ,
\end{equation}
which is in part small due to the nucleon mass suppression in Eq.~\eqref{eq:Ki}. The asymmetry only contains quadratic couplings, which also appear in the $R$ correlation. The constraints on these couplings from EDM measurements and the $R$ coefficient range from $\mathcal{O}(10^{-3})-\mathcal{O}(10^{-6})$ \cite{Koz09, Hub03}. Improving these bounds in radiative $\beta$ decay would require a measurement of the asymmetry to better than $10^{-11}$.  

($ii$) At leading order, the interactions in Eq.~\eqref{LowELag} give
\begin{eqnarray}
K & = &-16\frac{c_w}{e g}  \frac{E_e}{k\cdot p_e}  (g_A^2 +  g_V^2) \textrm{Re}\;C_{\varphi \tilde{W} B} - 8 s_w \frac{1 }{k\cdot p_e}\frac{\sqrt{2}M_W}{ e g} g_A g_T \eta_{qW}\textrm{Im}( C_{Wd}^* +C_{Wu}) \notag \\
&& + \frac{s_w}{e g}\textrm{Im} C_{\varphi ud} \left(8  \frac{E_e}{k\cdot p_e} (g_A^2-g_V^2) + 4\frac{ 1}{\omega} (g_A^2+g_V^2)\right) \ .
\end{eqnarray}
For $\omega_{\textrm{min}}=0.3$ MeV, the asymmetry for neutron radiative $\bt$ decay is
\bea\label{eq:as}
\mathcal{A}& =& -2\times 10^{-11} \;\textrm{Re}\;C'_{\varphi \tilde{W} B}  + 2\times10^{-7}\, \textrm{Im}({C'}_{Wd}^* +  C'_{Wu})\nn \\ &+&  4\times 10^{-12}\;\textrm{Im} C'_{\varphi u d}  \ ,
\eea
in terms of the couplings at $\Lambda=1\, \text{TeV}$. We used $g_A=1.27$, $s_w^2=0.23$, $g=0.64$, $g_T\sim 1$ and $M_W=80.4$ GeV. For clarity we have redefined $C'\equiv v^2 C$, such that the couplings $C'$ are dimensionless. 
Clearly, the contribution of these operators to the asymmetry is rather small. However, the sensitivity of the asymmetry to the $T$-odd BSM sources can be improved by choosing isotopes with larger $Q$-values. For $^{37}$K \cite{behr}, we find, for example, that it is 20 times more sensitive than neutron decay.   
In the next Section we discuss the stringent limit from EDMs on these couplings.

\section{Constraints from EDMs}\label{sec:EDM}
\begin{figure}
	\centering
		\includegraphics[width=0.60\textwidth]{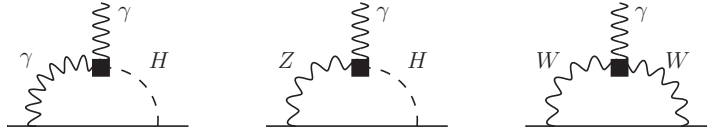}
	\caption{One-loop contributions of $Q_{\vp \tilde W B}$ to the quark and electron EDMs. }
	\label{fig:qedm}
\end{figure}
 The second class of operators also contribute to the neutron EDM (nEDM) and electron (eEDM). At the scale $M_W$, these operators induce
\bea
\vL_{\text{EDM}} &=&-\frac{i}{2} \sum_{f = u,d,e} d_f e Q_f m_f \, \bar \psi_f \simu\ga_5 \psi_f F_{\mu\nu}- i\, \text{Im}\, \Xi_1 \big[\bar u_R\ga^\mu d_R\, \bar d_L\ga_\mu u_L -\bar d_R\ga^\mu u_R\, \bar u_L\ga_\mu d_L\big]\nn\\
&& - i\, \text{Im}\, \Xi_8 \big[\bar u_R\ga^\mu t^a d_R\, \bar d_L\ga_\mu t^au_L -\bar d_R\ga^\mu t^au_R\, \bar u_L\ga_\mu t^ad_L\big],\label{LowELagEDM}
\eea
where $d_{u,d}$ represent the up- and down-quark EDM, $d_e$ is the electron EDM ($d_e^{\textrm{exp}}\equiv e m_e d_e$), and $\Xi_{1,8}$ are $CP$-odd four quark operators. The contributions from $C_{\vp ud}$, $C_{\vp \tilde WB}$, and $C_{qW}$ to the couplings in Eq.\ \eqref{LowELagEDM} are listed in Table \ref{tab:EDM}. 
 
\noindent $\boldsymbol{Q_{\vp ud}}$\\
Table \ref{tab:EDM} shows that, at leading order $C_{\vp ud}$, only contributes to $\Xi_1$. This interaction is generated after integrating out the $W^\pm$ boson through a tree-level diagram. The relevant interaction is similar to Fig.\ \ref{fig:2} without the photon, and where the $W^\pm$ should now be coupled to quarks instead of leptons \cite{Ng12}. As $Q_{\vp ud}$ does not evolve under QCD renormalization, the relation in Table \ref{tab:EDM} is independent of $\Lambda$.

\begin{table}
	\centering
\begin{tabular}{  c || c cc}
\hline\hline
 & $C_{\vp ud }(\Lambda)$ & $\qquad\frac{1}{gg'}\text{Re}\,C_{\vp \tilde WB }(\Lambda = 1\, (10)\, \text{TeV})$ & $\qquad\frac{\sqrt{2} v s_w}{e m_q Q_q}\text{Im}\, C_{qW}(\Lambda = 1\, (10)\, \text{TeV})$ \\
   \hline	\hline		
$ d_u(M_W)$ &$-$ & $ -6.2\,(-11)\times 10^{-2}$ &$ -0.80 \,(-0.69)$ \\
$ d_d(M_W)$ &$-$ & $ -11\,(-19)\times 10^{-2}$ &$0.80\, (0.69) $ \\
$ d_e(M_W)$ &$-$ & $ -5.3\,(-10)\times 10^{-2}$ &$- $ \\
$\Xi_1(M_W)$ & 1 & $-$ &$-$\\
$\Xi_8(M_W)$ & $-$ & $-$ &$-$\\
  \hline  \hline
\end{tabular}
	\caption{\footnotesize{The couplings in Eq.\ \eqref{LowELagEDM} at the scale $M_W$ in terms of the BSM couplings $C_{\vp ud}$, $C_{\vp \tilde WB}$, and $C_{qW}$ at the scale $\Lambda = 1\, (10)\, \text{TeV}$. The row headings are given by the column headings multiplied by the corresponding table entry, a dash indicates there is no contribution at leading order in perturbation theory. For more details, see  Ref.\ \cite{Dekens:2013zca}. }}
	\label{tab:EDM}
\end{table} 

\noindent$\boldsymbol{Q_{\vp \tilde W B}}$\\
In contrast, $Q_{\vp \tilde W B}$ does not contribute to the interactions in Eq.\ \eqref{LowELagEDM} at the tree-level, but it induces quark EDMs at the one-loop-level. This operator also contains $\textrm{Higgs}-\ga\ga$ and $\textrm{Higgs}-Z\ga$ interactions. These interactions and the $WW\gamma$ interaction in Eq.~\eqref{HighELag} contribute to the quark EDMs through the diagrams in Fig.\ \ref{fig:qedm} \cite{DeRujula:1990db, Dekens:2013zca}. As a consequence, the operator $Q_{\vp \tilde WB}$ mixes with the quark EDMs when it is evolved from the scale of new physics, $\Lambda$, down to $M_W$. The electron EDM (eEDM) is induced through the same mechanism. The results in Table \ref{tab:EDM} take into account both the mixing between $C_{\vp  \tilde WB}$ and $d_q$, and the running of $d_q$\footnote{Like $Q_{\vp ud}$, the operator $Q_{\vp \tilde WB}$ does not evolve under one-loop QCD renormalization.} \cite{Ciuchini:1993fk,Degrassi:2005zd}. 

\noindent$\boldsymbol{Q_{uW}$ \bf and $Q_{dW}}$\\
Finally, the $Q_{qW}$ operators contribute to $d_{u, d}$ directly, and we have, 
\bea
d_u(\Lambda) = -\frac{\sqrt{2}v s_w}{e Q_u m_u }\text{Im}\,C_{uW}(\Lambda),\qquad d_d(\Lambda) = \frac{\sqrt{2}v s_w}{e Q_d m_d }\text{Im}\,C_{dW}(\Lambda).
\eea
After taking into account the running of the quark EDMs, we obtain the results in Table \ref{tab:EDM}.
\newline

\begin{table}
	\centering
\begin{tabular}{  c || c cc}
\hline\hline
nEDM & $\textrm{Im}\;C'_{\vp ud }(\Lambda)$ & $\textrm{Re}\; C'_{\vp \tilde WB }(\Lambda)$ & $\textrm{Im}\;C'_{qW}(\Lambda)$ \\
   \hline	\hline		
$ \Lambda = 1\,  \text{TeV}$ &$ 1.0\times 10^{-5}  $&$ 1.8\times 10^{-4} $& $2.1\times 10^{-10}$ \\
$ \Lambda = 10\,  \text{TeV}$ &$1.0\times 10^{-5}$ & $ 1.0\times 10^{-4}$ &$ 2.2\times 10^{-10}$ \\
\end{tabular}\qquad
\begin{tabular}{  c || c }
\hline\hline
eEDM&  $\textrm{Re}\; C'_{\vp \tilde WB }(\Lambda)$\\\hline\hline
 $\Lambda = 1\, \,\text{TeV}$ &$2.3\times 10^{-6}$ \\
  $\Lambda = 10\,\text{TeV}$ &$1.2\times 10^{-6}$ \\
 \end{tabular}
	\caption{\footnotesize{Bounds on the couplings of the dimension-six operators 	in Table \ref{tab:ops}  due to the limits on the neutron and electron EDM. Again $C'\equiv v^2C$ to make the primed couplings dimensionless. The bounds are shown for two values of the scale of new physics, $\Lambda =1,\, 10\,  \text{TeV}$. In the cases that an operator contributes in several ways to $d_n$ we present the strongest bound, not taking into account possible cancellations. Only the $Q_{\vp \tilde WB}$ gives rise to a significant eEDM.}}
	\label{tab:bounds}
\end{table} 
 \begin{table}
	\centering
\
\end{table} 

The induced interactions at the scale $M_W$ have to be evolved to the low energies where EDM experiments take place. The renormalization group equations (RGEs) for the quark EDMs and the four-quark operators give \cite{Ciuchini:1993fk,Degrassi:2005zd, An10, Hisano:2012cc,Dekens:2013zca} 
\bea
&d_q(\MQCD) =0.48 \, d_q(M_W) , &\nn\\
&\text{Im}\, \Xi_1(\MQCD) = 1.1\,\Xi(M_W) ,\qquad \text{Im}\, \Xi_8(\MQCD) =1.4\, \Xi(M_W)  , 
\label{qEDMrunning}\eea 
where $\MQCD\approx 1\, \text{GeV}$ is the QCD scale, while the eEDM does not evolve under one-loop QCD renormalization. For calculation of the nEDM in terms of $d_q$ and $\Xi_{1,8}$ we use the following naive dimensional analysis \cite{Manohar:1983md,Weinberg:1989dx} (NDA) estimates,
\bea
d_n^{d_{q}} = \Or(e Q_q m_q)\,d_{u,d}(\MQCD),\qquad d_n^{\Xi} =  \Or \bigg(\frac{e \MQCD}{(4\pi)^2}\bigg) \text{Im}\, \Xi_{1,8}(\MQCD).\label{dn}
\eea
The estimate for $d_n^{d_{q}}$ is in agreement with QCD sum-rule results \cite{Pospelov:2000bw,Pospelov:2005pr}, while the estimate of $d_n^\Xi$ agrees with the results of Refs.~\cite{deVries:2012ab,Seng:2014pba,Maiezza:2014ala}.
%these analyses also indicate that the larger estimates used in Refs.\ \cite{Ng12,An:2009zh} are an overestimation. 
Combining Table \ref{tab:EDM}, Eq.\ \eqref{qEDMrunning} and \eqref{dn} with the upper limit on the nEDM, $|d_n|\leq 2.9 \times 10^{-26} e\, \text{cm}$ \cite{Baker:2006ts}, and eEDM, $|d_e^{\textrm{exp}}|\leq 8.7\times 10^{-29} e\,\text{cm}$ \cite{Baron:2013eja}, we finally obtain the bounds shown in Table \ref{tab:bounds}. 

\section{Conclusion}\label{sec:Conclusion}
Radiative $\beta$ decay offers the possibility to study a \textit{spin-independent} $T$-violating triple-correlation coefficient $K$. We have considered $T$-violating BSM physics arising above the electroweak scale that contributes to this correlation. The dimension-6 operators that contribute to $K$ also contribute to the spin-dependent EDMs. The EDM limits therefore stringently constrain these operators. In fact, comparing the EDM bounds in Table \ref{tab:EDM} to Eq.\ \eqref{eq:as}, we find that improving the EDM bounds would require a measurement of the neutron asymmetry better than $10^{-16}$. This accuracy cannot be reached in present experiments. In conclusion, the $T$-odd correlation $K$ is not "EDM-safe" when considering $CP$-violating dimension-6 operators above the electroweak scale. 

\begin{acknowledgments}
This research grew out of the {\em 33$^{\,rd}$ Solvay Workshop on Beta Decay Weak Interaction Studies in the Era of the 
LHC} (Brussels, September 3-5, 2014). We would like to thank Dani\"el Boer, Jordy de Vries and Rob Timmermans for helpful discussions.
We thank John Behr for information on the experimental prospects. 
This research was supported by the Dutch Stichting voor Fundamenteel Onderzoek der Materie
(FOM) under Programmes 104 and 114. \\
\end{acknowledgments}


\begin{thebibliography}{99}

\bibitem{Sak67} A. Sakharov, Pisma Zh. Eksp. Teor. Fiz. {\bf 5}, 32 (1967).
\bibitem{Riotto:1999yt} 
  A.~Riotto and M.~Trodden,
  Ann.\ Rev.\ Nucl.\ Part.\ Sci.\  {\bf 49}, 35 (1999).

\bibitem{Kuzmin} 
  V.~A.~Kuzmin, M.~E.~Shaposhnikov and I.~I.~Tkachev,
  Phys.\ Rev.\ D {\bf 45}, 466 (1992).
\bibitem{Jac57} J. Jackson, S. Treiman and H. Wyld, Phys. Rev. {\bf 106}, 517 (1957).
\bibitem{Ng12} J. Ng and S. Tulin, Phys. Rev. D {\bf 85}, 033001 (2012).
\bibitem{Her05} P. Herzceg, J. Res. Natl. Inst. Stand. Technol. {\bf 110}, 453 (2005). 
\bibitem{Mum11}
  H.~P.~Mumm {\it et al.},
  %``A New Limit on Time-Reversal Violation in Beta Decay,''
  Phys.\ Rev.\ Lett.\  {\bf 107},  102301 (2011).
\bibitem{Chu12}
  T.~E.~Chupp {\it et al.},
  %``Search for a T-odd, P-even Triple Correlation in Neutron Decay,''
  Phys.\ Rev.\ C {\bf 86},  035505 (2012).


\bibitem{Vos15} K. K. Vos {\it et al.}, in preparation.
\bibitem{Khr91} I. Khriplovich, Nucl. Phys. B {\bf 352}, 385 (1991).
\bibitem{Koz09}
  A.~Kozela, G.~Ban, A.~Bia\l{}ek, K.~Bodek, P.~Gorel, K.~Kirch, S.~Kistryn and M.~Ku\'zniak {\it et al.},
  %``Measurement of the Transverse Polarization of Electrons Emitted in Free Neutron Decay,''
  Phys.\ Rev.\ Lett.\  {\bf 102},  172301 (2009).
	\bibitem{Hub03}
  R.~Huber, J.~Lang, S.~Navert, J.~Sromicki, K.~Bodek, S.~Kistryn, J.~Zejma and O.~Naviliat-Cuncic {\it et al.},
  %``Search for time reversal violation in the beta decay of polarized Li-8 nuclei,''
  Phys.\ Rev.\ Lett.\  {\bf 90},  202301 (2003).

\bibitem{Bra01}
  V.~V.~Braguta, A.~A.~Likhoded and A.~E.~Chalov,
  %``T odd correlation in the K(l3 gamma) decay,''
  Phys.\ Rev.\ D {\bf 65}, 054038 (2002) 
   [Phys.\ Atom.\ Nucl.\  {\bf 65}, 1868 (2002)]
   [Yad.\ Fiz.\  {\bf 65}, 1920 (2002)].
\bibitem{Gar12} S. Gardner and D. He, Phys. Rev. D { \bf 86}, 016003 (2012).
\bibitem{Gar13} S. Gardner and D. He, Phys. Rev. D { \bf 87}, 116012 (2013).

\bibitem{Her97}
  P.~Herczeg and I.~B.~Khriplovich,
  %``Time reversal violation in Beta decay in the standard model,''
  Phys.\ Rev.\ D {\bf 56},  80 (1997).
\bibitem{Gar13a} S. Gardner and D. He, Hyperfine Interact. 214, 71 (2013). 



\bibitem{Grz10} 
  B.~Grzadkowski, M.~Iskrzy\'nski, M.~Misiak and J.~Rosiek,
  %``Dimension-Six Terms in the Standard Model Lagrangian,''
  JHEP {\bf 1010}, 085 (2010).


\bibitem{Cir10} V. Cirigliano, J. P. Jenkins, M. Gonz\'ales-Alonso, Nucl. Phys. B {\bf 830}, 95 (2010).
\bibitem{Her01} P. Herczeg, Prog. Part. Nucl. Phys. {\bf 46}, 413 (2001).
\bibitem{Lee56} 
  T.~D.~Lee and C.~N.~Yang,
  %``Question of Parity Conservation in Weak Interactions,''
  Phys.\ Rev.\  {\bf 104}, 254 (1956).


\bibitem{Ciuchini:1993fk} 
  M.~Ciuchini, E.~Franco, L.~Reina and L.~Silvestrini,
  Nucl.\ Phys.\ B {\bf 421}, 41 (1994).
\bibitem{Degrassi:2005zd} 
  G.~Degrassi, E.~Franco, S.~Marchetti and L.~Silvestrini,
  JHEP {\bf 0511}, 044 (2005).
\bibitem{Ber04} V. Bernard, S. Gardner, U.-G. Mei\ss ner, and C. Zhang, Phys. Lett. B {\bf 593}, 105 (2004) [Erratum-ibid. \textbf{599}, 438 (2004)]. 
\bibitem{behr} J. Behr, private communication. 
\bibitem{DeRujula:1990db} 
  A.~De R\'ujula, M.~B.~Gavela, O.~P\`ene, and F.~J.~Vegas,
  Nucl.\ Phys.\ B {\bf 357}, 311 (1991).
\bibitem{Dekens:2013zca} 
  W.~Dekens and J.~de Vries,
  JHEP {\bf 1305}, 149 (2013).




  
	\bibitem{An10}
	H. An, X. Ji, and F. Xu, JHEP {\bf 1002}, 043 (2010).
	
  \bibitem{Hisano:2012cc} 
  J.~Hisano, K.~Tsumura, and M.~J.~S.~Yang,
  Phys.\ Lett.\ B {\bf 713}, 473 (2012).  


\bibitem{Manohar:1983md} 
 A.~Manohar and H.~Georgi,
  Nucl.\ Phys.\ B {\bf 234}, 189 (1984).
 
\bibitem{Weinberg:1989dx} 
  S.~Weinberg,
  Phys.\ Rev.\ Lett.\  {\bf 63}, 2333 (1989).  
  
\bibitem{Pospelov:2000bw} 
M.~Pospelov and A.~Ritz,
Phys.\ Rev.\ D {\bf 63}, 073015 (2001).

\bibitem{Pospelov:2005pr} 
 M.~Pospelov and A.~Ritz,
 Ann. Phys.\  {\bf 318}, 119 (2005).

\bibitem{deVries:2012ab} 
  J.~de Vries, E.~Mereghetti, R.~G.~E.~Timmermans, and U.~van Kolck,
  Ann. Phys.\  {\bf 338}, 50 (2013).

\bibitem{Seng:2014pba} 
  C.~Y.~Seng, J.~de Vries, E.~Mereghetti, H.~H.~Patel, and M.~Ramsey-Musolf,
  Phys.\ Lett.\ B {\bf 736}, 147 (2014).
\bibitem{Maiezza:2014ala} 
  A.~Maiezza and M.~Nemev\v{s}ek,
  Phys.\ Rev.\ D {\bf 90}, 095002 (2014).


 
 \bibitem{Baker:2006ts} 
  C.~A.~Baker {\it et al.},
  Phys.\ Rev.\ Lett.\  {\bf 97}, 131801 (2006).
  
\bibitem{Baron:2013eja} 
  J.~Baron {\it et al.}  [ACME Collaboration],
  Science {\bf 343}, 269 (2014).

\end{thebibliography}
\end{document}